\documentclass{aa} 

\usepackage{txfonts}
\usepackage{graphicx}
\usepackage[english]{babel}
\usepackage{natbib}
\usepackage[pdftex]{hyperref}
\bibpunct{(}{)}{;}{a}{}{,}

\begin{document}

\title{Formation of the planet around the millisecond pulsar J1719--1438}

\author{
L.~M.~van~Haaften \inst{\ref{radboud}} \and
G.~Nelemans \inst{\ref{radboud},\ref{leuven}} \and
R.~Voss \inst{\ref{radboud}} \and
P.~G.~Jonker \inst{\ref{sron},\ref{radboud},\ref{cfa}}
}

\institute{
Department of Astrophysics/ IMAPP, Radboud University Nijmegen, P.O. Box 9010, 6500 GL Nijmegen, The Netherlands, \email{L.vanHaaften@astro.ru.nl} \label{radboud} \and
Institute for Astronomy, KU Leuven, Celestijnenlaan 200D, 3001 Leuven, Belgium \label{leuven} \and
SRON, Netherlands Institute for Space Research, Sorbonnelaan 2, 3584 CA, Utrecht, The Netherlands \label{sron} \and
Harvard-Smithsonian Center for Astrophysics, 60 Garden Street, Cambridge, MA 02138, USA \label{cfa}
}

\abstract{Recently the discovery of \object{PSR J1719--1438}, a $5.8$ ms pulsar with a companion in a $2.2$ hr orbit, was reported. The combination of this orbital period and the very low mass function is unique. The discoverers, Bailes et al., proposed an ultracompact X-ray binary (UCXB) as the progenitor system. However, the standard UCXB scenario would not produce this system as the time required to reach this orbital period exceeds the current estimate of the age of the Universe. The detached state of the system aggravates the problem. The inclination of the system is an important unknown, and Bailes et al. noted that for very low (a priori very unlikely) inclinations the system is better explained as having a brown dwarf companion rather than an UCXB origin.} 
{We want to understand the evolutionary history of PSR J1719--1438, and determine under which circumstances it could have evolved from an UCXB.} 
{We model UCXB evolution varying the donor size and investigate the effect of a wind mass loss from the donor, and compare the results with the observed characteristics of PSR J1719--1438.} 
{An UCXB can reach a $2.2$ hr orbit within the age of the Universe, provided that 1) the millisecond pulsar can significantly heat and expand the donor by pulsar irradiation, or 2) the system loses extra orbital angular momentum, e.g. via a fast wind from the donor.} 
{The most likely scenario for the formation of PSR J1719--1438 is UCXB evolution driven by angular momentum loss via the usual gravitational wave emission, which is enhanced by angular momentum loss via a donor wind of $\gtrsim 3 \times 10^{-13}\ M_{\odot}\mbox{yr}^{-1}$. Depending on the size of the donor during the evolution, the companion presently probably has a mass of $\sim \! 1-3$ Jupiter masses, making it a very low mass white dwarf as proposed by Bailes et al. Its composition can be either helium or carbon-oxygen. A helium white dwarf companion makes the long (for an UCXB) orbital period easier to explain, but the required inclination makes it a priori less likely than a carbon-oxygen white dwarf.} 

\keywords{stars: pulsars: individual: PSR J1719--1438 -- stars: binaries: close -- planets and satellites: formation}
\authorrunning{van~Haaften, Voss, Nelemans \& Jonker}
\titlerunning{Formation of the planet around PSR J1719--1438}

\maketitle

\section{Introduction} 

Even though the first exoplanets were discovered around a millisecond pulsar, \object{PSR B1257+12} \citep{wolszczan1992}, the discovery of a companion around \object{PSR J1719--1438} \citep[hereafter B11]{bailes2011} marked the first time a millisecond pulsar with a very low mass companion in a short orbit has been found. The orbital period is $2.177$ hr ($130.6$ min) and the mass function is $7.85 \times 10^{-10}\ M_{\odot}$, implying a minimum companion mass of $1.16 \times 10^{-3}\ M_{\odot}\ (1.47 \times 10^{-3}\ M_{\odot})$ in the case of a $1.4\ M_{\odot}\ (2\ M_{\odot})$ neutron star. The system is detached and there is no information on the chemical composition of the companion. The minimum mean density of the companion is $23.3$ g cm$^{-3}$ as follows from the relation between the mean density of the companion Roche lobe and the orbital period.

B11 stated that the high companion density, the orbital period, and the $5.8$ ms spin period of the pulsar are consistent with a history as an ultracompact X-ray binary (UCXB), which is a binary consisting of a white dwarf (like) donor transferring mass to a neutron star, forced by angular momentum loss via gravitational wave emission \citep{verbunt1995chapterlewin}.

B11 themselves noted the result by \citet{deloye2003} that an UCXB reaches an orbital period of $\sim \! 90$ min after $5-10$ Gyr. The discrepancy between $90$ min and $130.6$ min is very significant, however, because the orbital period increases slowly at longer orbital periods.
Furthermore, B11 did not satisfactorily answer why the system has become detached. They did suggest a change in the exponent of the white dwarf mass-radius relation near the present mass (assuming a near edge-on orbit) as a natural cause of detachment, but such a change merely changes the mass transfer rate and cannot lead to detachment.

In this paper, we will investigate two modifications to the UCXB evolution that could resolve these issues. These are 1) a larger donor radius and 2) additional angular momentum loss, apart from that driven by gravitational wave radiation. The actual scenario may be a combination of these.
In Sect. \ref{problems} we explain the problems, and in Sect. \ref{resolutions} we present possible resolutions.

\section{Problems with the standard UCXB scenario}
\label{problems}

Four intrinsic properties of PSR J1719--1438 are known: the orbital period, the maximum age of the system (the age of the Universe, which we take as $13.75$ Gyr, \citealp{jarosik2011}), the approximate primary mass as it is a neutron star, and the radial velocity of the primary. From these, two derived properties follow: the mass function, which gives an a priori likely companion mass, and the gravitational wave timescale $\tau_\mathrm{GW} = -(J_\mathrm{orb}/\dot{J}_\mathrm{orb})_\mathrm{GW}$ with $J_\mathrm{orb}$ the orbital angular momentum. The combination of the values of these parameters is inconsistent with the canonical UCXB scenario (e.g. \citealp{deloye2003,vanhaaften2012}) for the following reasons:

\begin{enumerate}

\item \textbf{Present evolutionary timescale} Even without making assumptions about the companion type, the implied companion mass is inconsistent with an evolution driven by angular momentum loss via gravitational wave emission, as shown in Fig. \ref{fig:gwrtimescale}, where the gravitational wave timescale is a function of the companion mass $M_\mathrm{c}$ and pulsar mass $M_\mathrm{p}$ via \citep{landau1975}
\begin{equation}
    \label{eqgwr}
    \tau_\mathrm{GW} = \frac{5}{32} \frac{c^{5}}{G^{5/3}} \frac{(M_\mathrm{p}+M_\mathrm{c})^{1/3}}{M_\mathrm{p}M_\mathrm{c}} \left( \frac{P_\mathrm{orb}}{2\pi} \right)^{8/3},
\end{equation}
where $P_\mathrm{orb}$ is the orbital period, $c$ the speed of light, $G$ the gravitational constant, and $M_\mathrm{c}$ is a function of the binary inclination $i$ given the measured mass function $(M_\mathrm{c} \sin i)^{3}/(M_\mathrm{p}+M_\mathrm{c})^{2}$ and an assumed pulsar mass. For normal (close to edge-on) inclinations, the gravitational wave timescale exceeds the age of the Universe by a large factor ($\sim \! 10-30$). The ratio between the gravitational wave timescale and the age of the Universe needs to be less than or close to $1$, if evolution is driven by gravitational wave emission, but the corresponding inclinations have a priori probabilities near $0.1\%$.

\begin{figure}
\resizebox{\hsize}{!}{\includegraphics{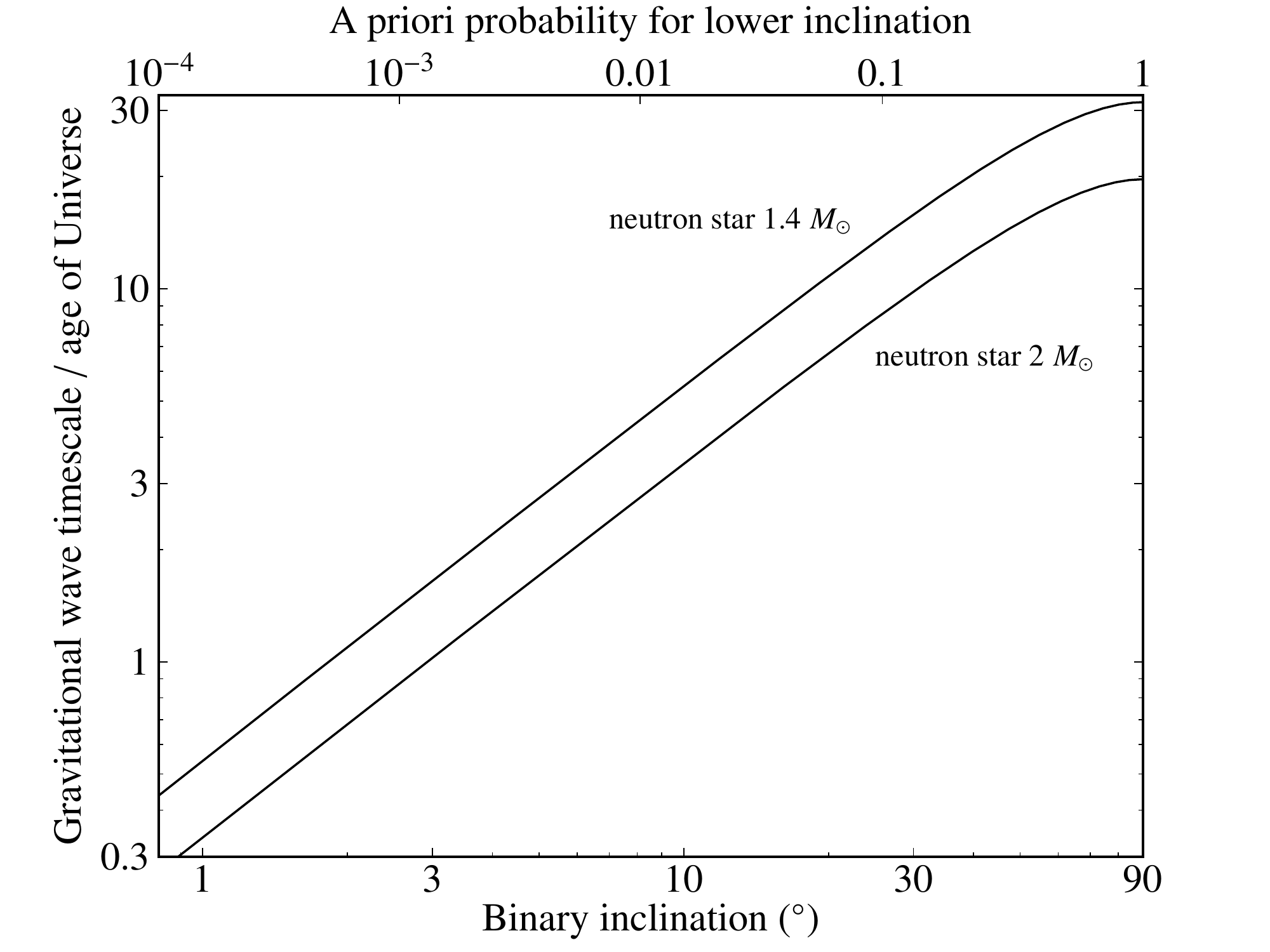}} 
\caption{Timescale for angular momentum loss via gravitational wave emission as a function of binary inclination for two assumed pulsar masses, for \object{PSR J1719--1438} with mass function $7.85 \times 10^{-10}\ M_{\odot}$. The upper horizontal axis gives the a priori probability for an inclination lower than indicated on the lower horizontal axis.}
\label{fig:gwrtimescale}
\end{figure}

\item \textbf{Orbital period} If we hypothesize PSR J1719--1438 to have originated from an UCXB, we can explore the previous argument in more detail. The orbital period of an UCXB increases during its evolution,\footnote{The orbital period increases because the exponent $\zeta_\mathrm{d}$ of the mass-radius relation of the white dwarf donor is lower than $1/3$, corresponding to a decreasing average donor density. This follows from the Roche lobe geometry and Kepler's third law.} at a rate determined by angular momentum loss via gravitational wave radiation. Figure \ref{fig:mp} (solid tracks) shows that an UCXB containing a donor close to zero-temperature radius (i.e., the radius of a white dwarf lacking support by thermal pressure) cannot reach the orbital period of PSR J1719--1438 of $130.6$ min within the age of the Universe. Donors that are heated have a larger radius, which results in a longer orbital period at a given age. To gain insight in how a larger radius translates into a longer orbital period, we simply parameterize a bloated donor by multiplying the radius by a fixed factor for all masses (dashed tracks). A factor $2$ is required for a helium white dwarf donor UCXB to reach an orbital period of $130.6$ min within the age of the Universe, and an even larger factor of $2.5$ for a carbon-oxygen white dwarf donor.

\begin{figure}
\resizebox{\hsize}{!}{\includegraphics{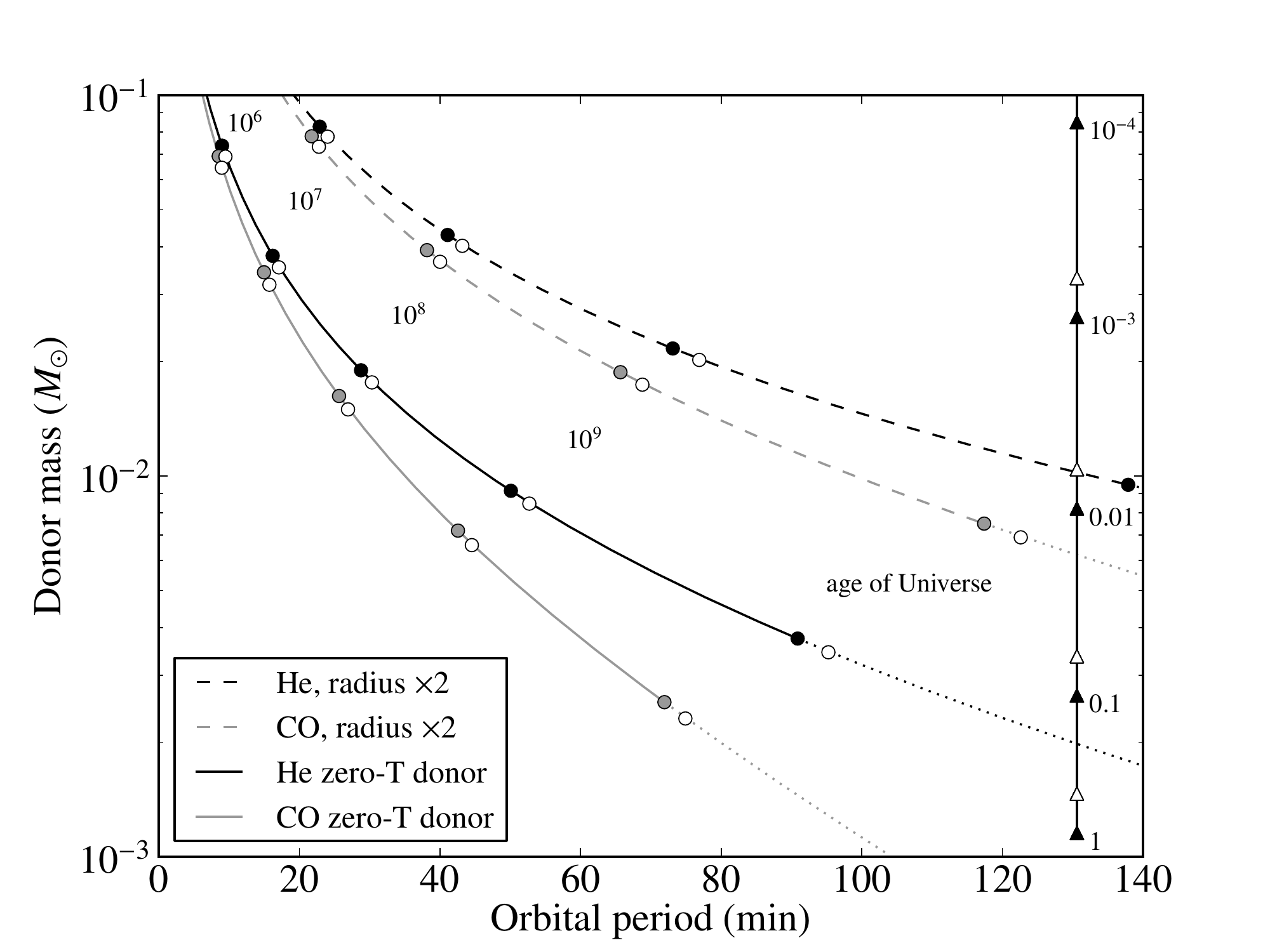}} 
\caption{Donor mass versus orbital period for UCXBs. The solid curves represent a zero-temperature carbon-oxygen (gray) and helium (black) white dwarf donor. The dashed curves are different from the solid curves only in that they have a twice as large donor radius at all masses. The circles on top of the solid and dashed curves mark the age of the system since the onset of mass transfer, the numbers associated with the circles indicate the ages in yr. Filled (open) symbols indicate an initial accretor mass of $1.4\ M_{\odot}\ (2\ M_{\odot})$. The dotted extensions to the curves go beyond the point that a UCXB with a $1.4\ M_{\odot}$ accretor can reach within the age of the Universe. The vertical solid line shows the present orbital period of PSR J1719--1438. The overlaying triangles give the a priori probabilities for the donor mass being higher than indicated, based on the mass function.}
\label{fig:mp}
\end{figure}

The onset of mass transfer is expected to have occurred several gigayears after the Big Bang, so less time is available for the described evolution, aggravating the problem. Moreover, it is possible that the system became detached long ago.

\item \textbf{Detached state} Since an UCXB at all times continues to lose angular momentum via gravitational waves, the donor will keep filling its Roche lobe and therefore is not expected to become detached. B11 suggested that an UCXB becomes detached when the mass-radius relation of the donor becomes $\sim \! 0$, but this is not the case. The behavior of the system does not qualitatively depend on the value or the sign of the exponent $\zeta_\mathrm{d}$ of the mass-radius relation of the donor $R_\mathrm{d} \propto M_\mathrm{d}^{\zeta_\mathrm{d}}$, as long as $\zeta_\mathrm{d}$ is not close to the exponent $\zeta_\mathrm{L} \approx -5/3$ of the donor mass-Roche lobe radius relation $R_\mathrm{L} \propto M_\mathrm{d}^{\zeta_\mathrm{L}}$. Here $M_\mathrm{d}, R_\mathrm{d}$ and $R_\mathrm{L}$ are the donor mass and the donor and Roche-lobe radii, respectively. The increasing exponent of the mass-radius relation of the donor resulting from its decreasing mass merely lowers the mass transfer rate (e.g. Eq. (6) in \citealt{savonije1986})\footnote{The simultaneously increasing gravitational wave timescale generally has a much larger impact on the mass transfer rate.}

\begin{equation}
    \label{mdot}
    -\dot{M}_\mathrm{d} = \frac{2}{\zeta_\mathrm{d} - \zeta_\mathrm{L}} \frac{M_\mathrm{d}}{\tau_\mathrm{GW}}
\end{equation}
and by itself cannot cause detachment, even if this exponent becomes zero or positive (which happens at the donor masses corresponding to the maxima of the dashed mass-radius curves in Fig. \ref{fig:mr0}).
A detached state suggests that the companion has shrunk by itself on a timescale shorter than the (at this point very long) evolutionary timescale.

\end{enumerate}

\section{Possible resolutions}
\label{resolutions}

\begin{figure}
\resizebox{\hsize}{!}{\includegraphics{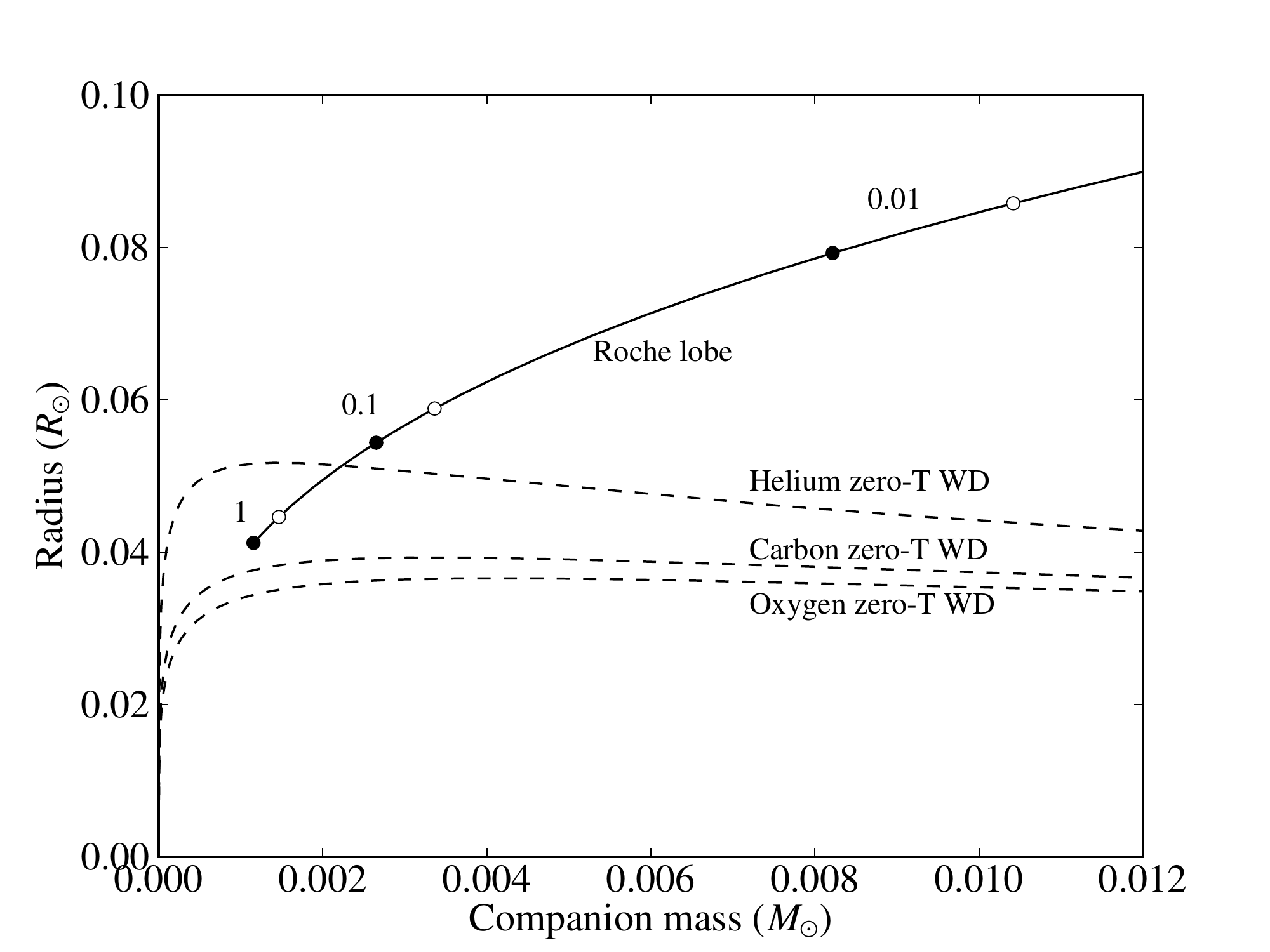}} 
\caption{Zero-temperature white dwarf mass-radius relations by Eggleton \citep{rappaport1987} for pure helium, carbon and oxygen compositions (dashed), and the present Roche-lobe radius against mass for the companion of \object{PSR J1719--1438}. Filled circles (neutron star mass of $1.4\ M_{\odot}$) and open circles (neutron star mass of $2\ M_{\odot}$) from low to high companion mass indicate the $1$, $0.1$ and $0.01$ a priori probabilities for a companion mass higher than indicated.}
\label{fig:mr0}
\end{figure}

Figure \ref{fig:mr0}, which is similar to the figure in B11, shows the range of allowed companion radii as function of its mass and composition. The zero-temperature radius (dashed) is the lower limit and the Roche-lobe radius (solid) the upper limit.
If the system is observed close to edge-on, a helium white dwarf is too large to fit inside the Roche lobe, while a carbon-oxygen white dwarf does fit in as long as it is not much larger than the zero-temperature radius. However, if we observe the system at an inclination of less than $31^{\circ}\ (41^{\circ})$ (a priori probability $14\%\ (24\%)$) in the case of a $1.4\ M_{\odot}\ (2\ M_{\odot})$ neutron star, the companion mass is sufficiently high for even a zero-temperature helium white dwarf to fit in its Roche lobe. If bloated, a lower inclination is required. Hydrogen-dominated planets or low-mass brown dwarfs have a much higher radius ($\sim \! 0.1\ R_{\odot}$) and therefore require a very low inclination.

\subsection{Bloated donor scenario}

The relatively long orbital period may have been reached within the age of the Universe if the donor has been bloated during a significant part of its lifetime. Figure \ref{fig:mp} shows that the present companion mass in this scenario is at least $\sim \! 0.01\ M_{\odot}$, which requires that we observe the system at an a priori unlikely inclination of less than $\sim \! 6.6^{\circ}$ ($0.7\%$), assuming a neutron star mass of $1.4\ M_{\odot}$. Alternatively, the $\sim \! 0.01\ M_{\odot}$ donor has subsequently lost a large amount of mass at an almost constant orbital period near the present orbital period, arriving at a mass of $\sim \! 10^{-3}\ M_{\odot}$. Before this mass loss event, the gravitational wave timescale would have been much shorter and consistent with its maximum age (the age of the Universe), see Fig. \ref{fig:gwrtimescale}.

Investigation of the mass-radius relation by \citet{deloye2003} showed that bloated factors of $2$ or higher, that would be necessary to explain the current observed properties of \object{J1719--1438}, are unlikely. So the bloated donor scenario can at most be a partial explanation.

\subsection{Additional angular momentum loss scenario}

If gravitational wave radiation is not the only mechanism for angular momentum loss, the real evolutionary timescale is shorter, or has been shorter in the past.
Empirical evidence from Cataclysmic Variables suggests that the angular momentum loss in systems below the period gap is higher (by a factor of $\sim \! 2.5$) than expected from gravitational wave emission alone \citep{knigge2011}. Also, \object{SAX J1808.4--3658}, a millisecond pulsar accreting from what is probably a $\sim \! 0.05\ M_{\odot}$ brown dwarf companion \citep{bildsten2001}, and therefore rather similar to an UCXB progenitor of \object{J1719--1438}, is losing more angular momentum than expected from gravitational wave radiation \citep{disalvo2008}.

If the difference between the exponents of the mass-radius relations of the donor and its Roche lobe becomes smaller, mass transfer is accelerated. This requires either less tendency of the Roche lobe to expand upon mass transfer (i.e., a higher (less negative) value of $\zeta_\mathrm{L}$), or a stronger expansion of the donor (a lower value of $\zeta_\mathrm{d}$). The latter is less likely to happen since both cold and heated white dwarfs tend to either weakly expand or shrink at low mass.\footnote{If the white dwarf core temperature remains constant due to tidal heating, the exponent of the mass-radius relation may diverge, leading to a dynamical instability that disrupts the companion \citep{bildsten2002}.} The former is more likely; a higher exponent $\zeta_\mathrm{L}$ of the donor mass-Roche lobe radius relation can be the result of mass loss from the system because this mass carries angular momentum, and therefore the semi-major axis will increase less upon mass transfer.

\subsubsection{Donor wind}

Mass lost directly from the donor in a fast wind (the Jeans mode) carries a large amount of specific angular momentum, because of the high mass ratio. The specific angular momentum of the donor relative to the orbit is equal to $M_\mathrm{a}/M_\mathrm{d}$, where $M_\mathrm{a}$ is the accretor mass.\footnote{At very high mass ratio the donor even absolutely carries almost all of the orbital angular momentum of the system in its orbit around the center of mass. Interestingly, in J1719--1438 the spin angular momentum of the pulsar is significant at $\sim \! 1/6$ of the orbital angular momentum.}
This wind can be caused by high-energy radiation from the millisecond pulsar, such as X-rays, gamma-rays from magnetosphere-accretion disk interaction and, when accretion has stopped, an electron-positron wind \citep{kluzniak1988,ruderman1989,shaham1992}. Heating by the pulsar wind has been observed in the accretion-powered millisecond X-ray pulsars \object{SAX J1808.4--3658} \citep{burderi2009} and \object{IGR J00291+5934} \citep{jonker2008} in quiescence. Similarly, heating of the donor by the hot neutron star has been observed in \object{EXO 0748--676} in quiescence \citep{bassa2009}. \citet{ratti2012} found evidence for a wind driven off the donor in this system.

\begin{figure}
\resizebox{\hsize}{!}{\includegraphics{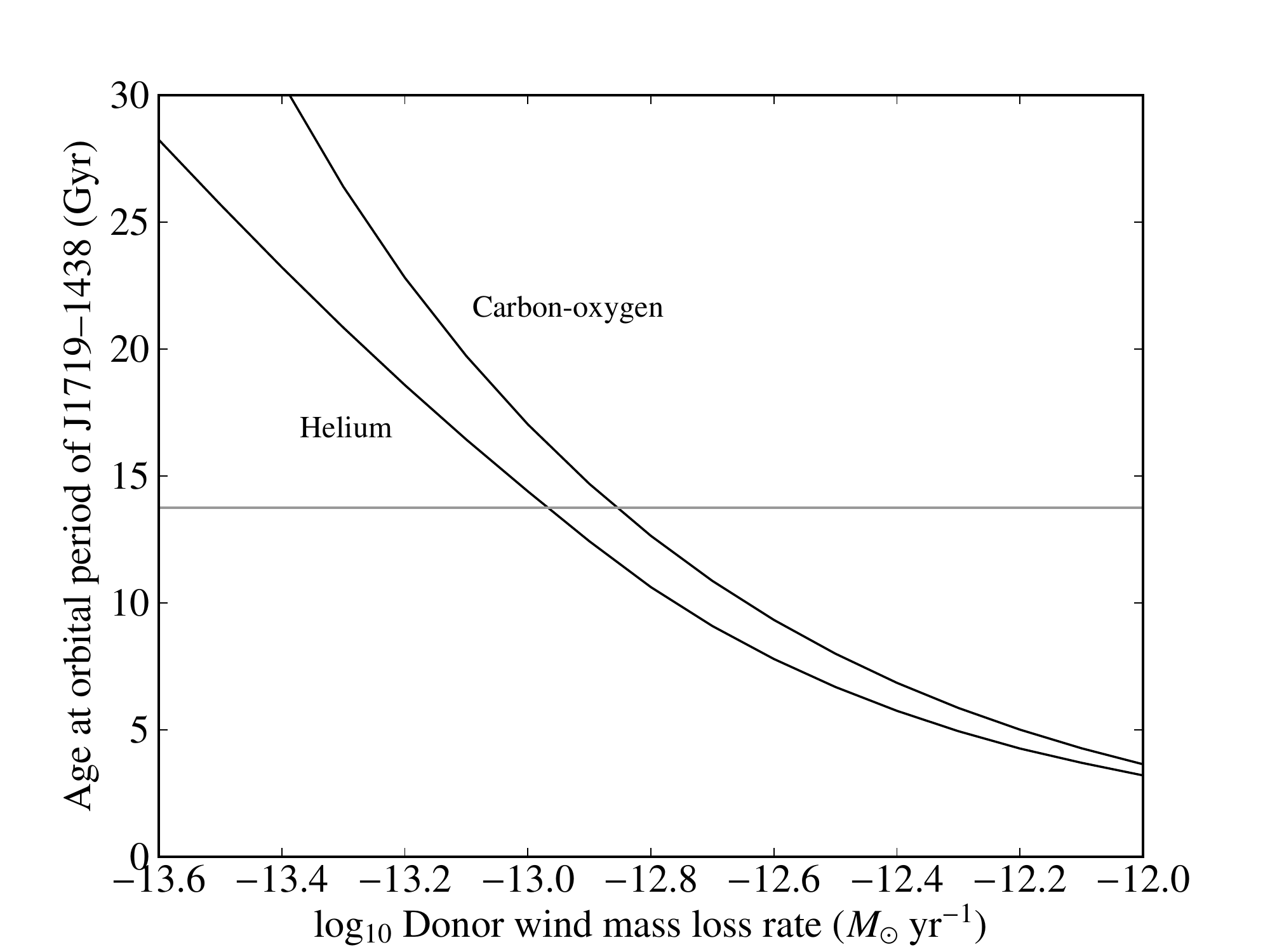}} 
\caption{Age of an UCXB with an initially $1.4\ M_{\odot}$ accretor when its orbital period equals the orbital period of \object{PSR J1719--1438} ($130.6$ min) versus donor mass loss rate, which is assumed to be fast and isotropic, and constant. The donor composition is indicated. The horizontal gray line gives the maximum age of PSR J1719--1438.} 
\label{fig:wind}
\end{figure}

Figure \ref{fig:wind} shows the effect the wind has on accelerating the evolution. When the wind mass loss rate is very low, reaching the orbital period of PSR J1719--1438 takes (much) longer than the age of the Universe, as shown in Sect. \ref{problems}. For a donor wind of $\sim \! 3 \times 10^{-13}\ M_{\odot}\mbox{yr}^{-1}$, however, evolution proceeds sufficiently fast to explain the observed orbital period. If the wind is assumed to commence later, e.g. below a threshold donor mass of $0.01\ M_{\odot}$, the required time increases by only $\sim \! 10-20\%$.

\subsection{Detachment at very low donor mass}

At the present mass of $\gtrsim 10^{-3}\ M_{\odot}$, the donor must have shrunk relatively rapidly to explain why it has become detached. The cause could be a change in radius not driven by mass loss, either due to steadily decreasing heating by the pulsar, which is expected given the declining accretion rate, pulsar spin period (at low donor mass, \citealp{vanhaaften2012}) and magnetic field strength, or due to changing thermal properties of the companion which could allow for more efficient cooling.

If the donor has non-degenerate outer layers due to heating from the pulsar, mass loss would actually shrink the donor, and if this happens rapidly, the remnant may become detached. This implies that the remnant must be quite close to the zero-temperature radius, and certainly much less bloated than the $\sim \! 0.01\ M_{\odot}$ object.

\subsection{Detachment due to thermal-viscous disk instability}

In particular in binaries with a low mass transfer rate, the accretion disk is subject to a thermal-viscous instability \citep{osaki1974,lasota2001} and periodically collapses. For a fast, isotropic wind from either accretor or donor, $a\, (M_\mathrm{a} + M_\mathrm{disk} + M_\mathrm{d}) = \mbox{constant}$, where $a$ is the semi-major axis and $M_\mathrm{disk}$ the disk mass, which is included because from an orbital dynamics perspective, the disk can be treated as belonging to the accretor. If we assume that during an outburst the entire disk is emptied, where almost all of the disk mass escapes the system, the orbit will expand by $\Delta a/a = M_\mathrm{disk}/(M_\mathrm{a} + M_\mathrm{d})$ (at very low donor mass $\Delta a \sim \! 1$ cm which means the donor does not actually detach but rather insufficiently overfills its Roche lobe to be able to prevent the orbit from shrinking).
We use the disk description by \citet{dunkel2006} for helium and carbon-oxygen composition to estimate the disk mass. At high mass ratio, the inner disk radius is taken equal to the speed-of-light cylinder radius, as described in \citet{vanhaaften2012}.

During the outburst, the donor detaches because its Roche lobe expands along with the orbit. The time it takes for the donor to re-attach follows from the orbital decay rate due to angular momentum loss (via gravitational wave radiation) $\dot{a} = 2a (\dot{J}/J)_\mathrm{orb}$.
The re-attachment time is at most $\sim \! 1$ yr, even for low viscosity (parameterized by $\alpha = 0.02$), low donor mass ($10^{-3}\ M_{\odot}$) and the loss of (almost) the entire disk mass from the system. The time it takes to rebuild the disk is of the order of $100$ yr \citep{vanhaaften2012}. 

The hypothesis that \object{PSR J1719--1438} is a system that at the present is only temporarily detached as part of a disk instability cycle would imply that less than a year before its discovery this system had a large outburst that caused the detached state, which would have made the system appear as a transient X-ray source. Also, the neutron star would still be very hot, and therefore bright in X-rays, as its cooling timescale is of the order of $10^{4}$ yr.

\section{Discussion and conclusions}
\label{conclusion}

In the a priori unlikely case the we observe the millisecond pulsar J1719--1438 nearly face-on, the companion could be a brown dwarf of $\sim \! 10-40$ Jupiter masses that is being evaporated and therefore has become detached from its Roche lobe. However, an optical non-detection of the system makes the presence of a relatively massive companion less likely (B11).

\object{PSR J1719--1438} is more plausibly explained by having an ultracompact X-ray binary progenitor. The system could have started as a regular UCXB of either helium or carbon-oxygen composition. Cooling due to heat emission and expansion may have caused the donor radius to eventually approach the zero-temperature radius, however, radiation from the pulsar can heat the outer layers of the donor. In particular at low density this effect can be significant. This heating can lead to a fast stellar wind from the donor which removes angular momentum from the system and accelerates the system's evolution, allowing longer orbital periods and lower companion masses than would be possible without such a wind. Moreover, the larger size of the donor also leads to a longer orbital period at a given age.

A combination of bloated donor, donor wind and low inclination can explain the properties of the system (the relatively long orbital period, why the present-day gravitational wave timescale most likely is much longer than the age of the Universe, and why the system is detached) without requiring an improbable contribution of either of these.

No excess dispersive delays have been found in the radio light curve (B11), so there is no observational evidence for ablation of the companion. But the required wind mass loss rate we find is much lower than a donor-evaporating black widow mass loss rate ($\sim \! 10^{-10}\ M_{\odot}\mbox{yr}^{-1}$, \citealp{burderi2009}) and therefore may not be observable.

Neither helium nor carbon-oxygen can be ruled out as the composition of the companion. A helium composition is a priori less likely since it requires a relatively special system inclination (the associated probability is less than $14\%$ in the case of a $1.4\ M_{\odot}$ neutron star) especially if it is bloated. However, since UCXBs with helium white dwarf donors have a longer orbital period than systems with carbon-oxygen white dwarfs, it takes less heating and angular momentum loss via a donor wind to explain the long (for UCXB standards) orbital period.

Limited feedback of angular momentum from the accretion disk to the orbit can in principle cause accelerated mass transfer over a prolonged period of time, but the occurrence of this process is unlikely \citep{priedhorsky1988,vanhaaften2012}.

The distance to \object{PSR J1719--1438}, $\sim \! 1.2$ kpc (B11), points to a Galactic Plane environment, so a formation involving dynamical interaction is not likely.

\begin{acknowledgements}
LMvH is supported by the Netherlands Organisation for Scientific Research (NWO). GN, RV and PGJ are supported by NWO VIDI grants.
\end{acknowledgements}

\bibliographystyle{aa}
\bibliography{lennart_refs}

\end{document}